\documentclass[journal]{IEEEtran}
\ifCLASSINFOpdf
\else
\fi
\usepackage{cite}
\usepackage{amsmath,amssymb,amsfonts}
\usepackage{algorithmic}
\usepackage{algorithm}
\usepackage{epsfig}
\usepackage{epstopdf}
\usepackage{graphicx}
\usepackage{textcomp}
\usepackage{xcolor}
\usepackage{lipsum}
\usepackage{stfloats}
\begin{document}
%
\title{The Property of Frequency Shift in 2D-FRFT Domain with Application to Image Encryption}
%

\author{Lei Gao,~\IEEEmembership{Member,~IEEE,}
        Lin Qi,~\IEEEmembership{Member,~IEEE,}
        Ling Guan,~\IEEEmembership{Fellow,~IEEE}
\thanks{L. Gao and L. Guan are with the Department of Electrical, Computer and Biomedical Engineering, Ryerson University, Toronto, ON M5B 2K3, Canada (email:iegaolei@gmail.com; lguan@ee.ryerson.ca).}
\thanks{L. Qi is with the School of Information Engineering, Zhengzhou University, Zhengzhou, China (email:ielqi@zzu.edu.cn).}}
\maketitle

\begin{abstract}
The Fractional Fourier Transform (FRFT) has been playing a unique and increasingly important role in signal and image processing. In this paper, we investigate the property of frequency shift in two-dimensional FRFT (2D-FRFT) domain. It is shown that the magnitude of image reconstruction from phase information is frequency shift-invariant in 2D-FRFT domain, enhancing the robustness of image encryption, an important multimedia security task. Experiments are conducted to demonstrate the effectiveness of this property against the frequency shift attack, improving the robustness of image encryption.
\end{abstract}
%
\IEEEpeerreviewmaketitle
\section{Introduction}
%
%
%
%
\IEEEPARstart{T}{he} Fourier Transform (FT) is one of the most important analysis tools used in physical optics and signal processing [1-3]. As a generalization of the FT, Fractional Fourier Transform (FRFT) was introduced in 1980 [4-5]. Different from the FT, the FRFT of a signal is flexibly operated at any angle with respect to the time axis on the time-frequency plane, generating a versatile representation for time-frequency distributions (TFDS) of the Cohen class. In fact, the conventional FT is a special case of the FRFT, when the operation angle is 90 degree with respect to the time axis. FRFT provides a powerful tool to analyze signals in the time-frequency domain [6]. The FRFT has since drawn the attention of researchers in the signal processing communities, and fractional operations have been introduced [7-8]. Typical examples include the fractional convolution [9], the fractional correlation [10-11], and the fractional filter [12], which extend the original operations.\\\indent In the Fourier representation of signals, a widely accepted confidence is that amplitude and phase tend to play different roles. Hayes [3] demonstrated that phase was more important than amplitude by reconstructing a multidimensional sequence from the phase part of its FT. In the past couple of decades, extensive related works also have been presented on FRFT. Signal reconstruction from amplitude or phase information of one dimensional FRFT (1D-FRFT) has been extensively investigated. In-depth analysis on amplitude and phase in FRFT domain was presented in [13-15]. It demonstrated that phase played more important roles than amplitude in FRFT domain [13]. Thus, phase retrieval using the FRFT was introduced for image encryption and examination of sensitivities of the various encryption keys [16]. In addition, the methods of multiple-parameter FRFT [17-18] were proposed and applied to the image feature extraction and representation.\\\indent To study the 2D time varying signals, extension of FRFT to two dimensions has also been conducted, in both continuous and discrete domains [19], laying foundations for further investigations in two-dimensional FRFT (2D-FRFT). The properties of spatial shift [20] and rotation invariance [21] in 2D-FRFT were investigated with applications to moving target detection and watermarking respectively. As a result, it is an urgent priority in investigating the characteristics of phase information and amplitude information in 2D-FRFT domain. Nevertheless, as a standing problem, frequency shift can introduce interference into the phase information, leading to poor performance on related applications [22-23]. As far as we know, studies on frequency shift in 2D-FRFT are limited.\\\indent To address the aforementioned issues, in this letter, we present a study of the properties of frequency shift in 2D-FRFT from amplitude and phase information with mathematical verification and computer simulations. The main contributions are summarized as follows.\\
\textbf{1}. It is demonstrated that the magnitude of image reconstruction from phase information is frequency shift-invariant in 2D-FRFT domain while the magnitude of reconstruction from amplitude information does not possess this property.\\
\textbf{2}. In application, we show that the utilization of this property improves robustness of image encryption.\\\indent The remainder of this letter is organized as follows: Section II reviews related work. Section III introduces and verifies the property of frequency shift in 2D-FRFT domain. Section IV presents application examples and Section V draws conclusions.
\section{Related Work}
In this section, we will briefly present the existing fundamentals of FRFT and 2D-FRFT, respectively.
\subsection{FRFT}
The transform of a 1D signal \emph{h}(\emph{t}) by FRFT is written as
\begin{equation}
{H_\alpha }\left( u \right) = \left\{ {{F_\alpha }\left[ {h\left( t \right)} \right]} \right\}\left( u \right) = \int_{ - \infty }^\infty  {h\left( t \right){K_\alpha }\left( {t,u} \right)dt,}
\end{equation}
with the transform kernel $ {{K_\alpha }\left( {t,u} \right)}$, in the following form
\begin{equation}
{K_\alpha }(t,u) =\left\{{\begin{array}{*{20}{c}}{{k_\alpha}\cdot \exp \left( {\begin{array}{*{20}{c}}
{i\frac{{{t^2} + {u^2}}}{2}\cot \alpha  - itu\csc \alpha}  \\
\end{array}} \right), \quad \alpha  \ne n\pi }  \\\
{\delta (t - u),\quad \quad \quad \quad \quad \quad \quad \quad \quad \quad \quad \quad {\rm{  }}\alpha  = 2n\pi }  \\\
{\delta (t + u),\quad \;\quad \quad \quad \quad \quad \quad \quad \quad {\rm{  }}\alpha  = (2n \pm 1)\pi }  \\
\end{array}} \right.
\end{equation}
where ${k_\alpha } = \sqrt {1 - i\cot \alpha /2\pi }$ $ (i = \sqrt {-1})$ and $ \alpha $ is the rotation angle in FRFT.
\subsection{2D-FRFT}
With two rotation angles $\alpha$ and $\beta$, 2D-FRFT provides two degrees of freedom coping with signal and image processing problems. Analytically, the definition of 2D-FRFT to a 2D signal \emph{d}(\emph{s}, \emph{t}) is given as
\begin{equation}
{D_{\alpha ,\beta }}(u,v) = \left\{ {{F_\beta }\left\{ {{F_\alpha }\left[ {d(s,t)} \right]} \right\}(u,t)} \right\}(u,v).
\end{equation}
Let the size of a discrete 2D signal \emph{g(p, q)} be (\emph{P}, \emph{Q}). The forward and inverse 2D-FRFT to a 2D discrete signal \emph{g(p,q)} are expressed as in [19]:
\begin{equation}
{G_{\alpha ,\beta}}(m,n) = \sum\limits_{p = 0}^{P - 1} {\sum\limits_{q = 0}^{Q - 1} {g(p,q)} } {K_{\alpha ,\beta }}(p,q,m,n),
\end{equation}
\begin{equation}
g(p,q) = \sum\limits_{m = 0}^{P - 1} {\sum\limits_{n = 0}^{Q - 1} {{G_{\alpha ,\beta}}(m,n)} } {K_{ - \alpha , - \beta }}(p,q,m,n),
\end{equation}
where ${K_{\alpha ,\beta}}(p,q,m,n)$ and ${K_{-\alpha ,-\beta}}(p,q,m,n)$ are the forward and inverse 2D discrete transform kernels, respectively.
\section{Frequency Shift in 2D-FRFT Domain}
\begin{figure*}[t]
\centering
\includegraphics[height=3.6in,width=6.0in]{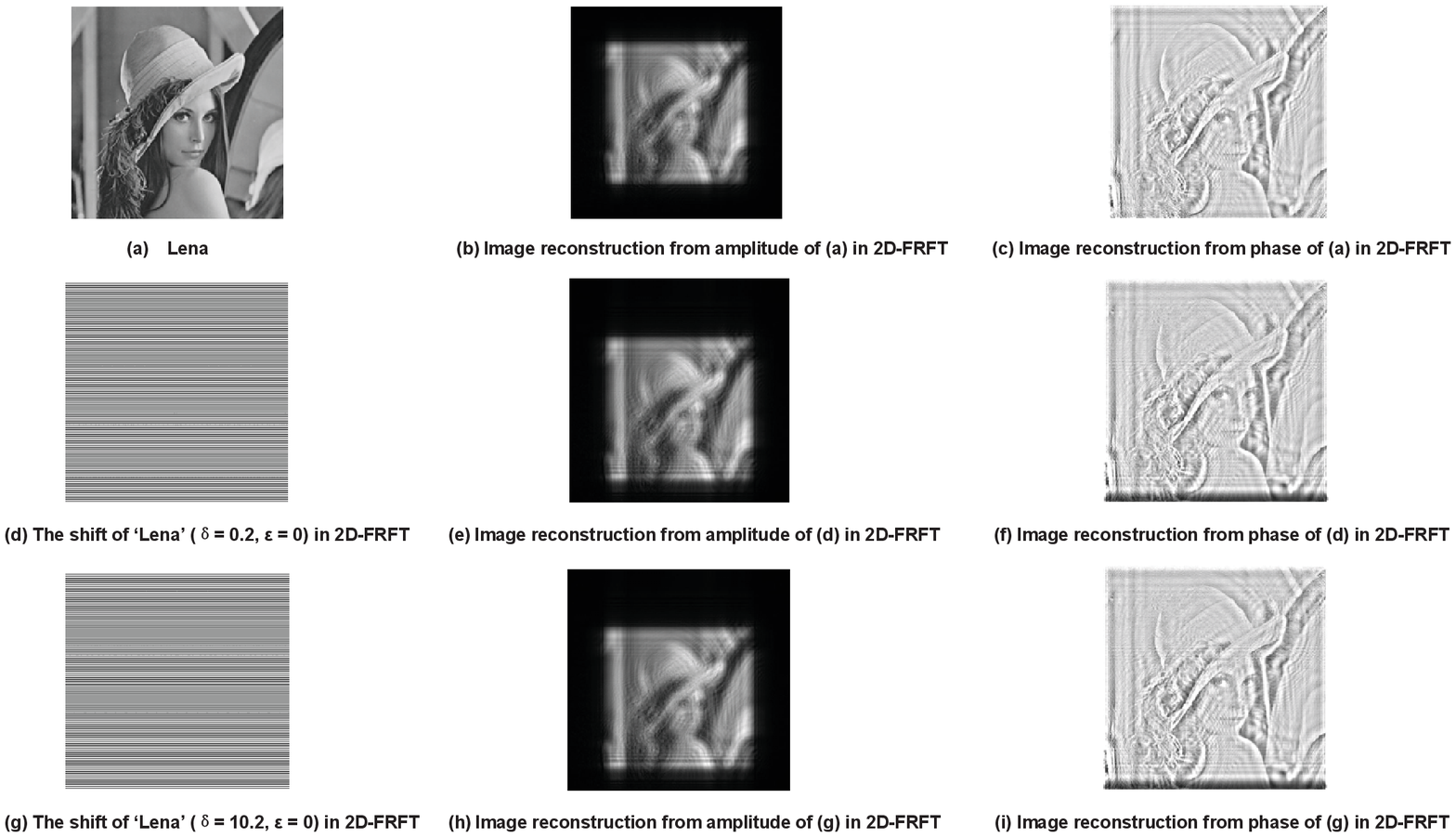}\\ Figure. 1 The simulation results on image `Lena'.\\
\end{figure*}
\subsection{Mathematical Derivation}
Again, the 2D-FRFT to a 2D signal \emph{f}(\emph{$x$}, \emph{$y$}) is expressed as
\begin{equation}
{F_{\alpha ,\beta }}(u,v) = \int\limits_{ - \infty }^{ + \infty } {\int\limits_{ - \infty }^{ + \infty } {{K_\alpha }(x,u)\cdot{K_\beta }(y,v)\cdot f(x,y)\cdot dxdy} },
\end{equation}
where $ {{K_\alpha }\left( {x,u} \right)}$ and $ {{K_\beta }\left( {y,v} \right)}$ are transform kernel functions defined in equation (2). Equation (6) is rewritten equivalently as follows
\begin{equation}
\begin{array}{l}
 {F_{\alpha ,\beta }}(u,v) = {A_{\alpha ,\beta }}(u,v)\cdot\exp (i2\pi {\varphi _\alpha }(u) + i2\pi {\varphi _\beta }(v)), \\
 \end{array}
\end{equation}
where ${A_{\alpha ,\beta }}(u,v)$ and $\exp (i2\pi {\varphi _\alpha }(u) + i2\pi {\varphi _\beta }(v))$ represent the amplitude and phase components of equation (6). \\\indent Then, the amplitude part $ {f_{A}(x,y)} $ and the phase part $ {f_{\varphi}(x,y)} $ in the space domain are reconstructed from equation (7) by inverse 2D-FRFT transform [19], and are defined in equations (8) and (9), respectively.
\begin{equation}
{f_{A}}(x,y) = {F_{ - \alpha , - \beta }}({A_{\alpha ,\beta }}(u,v)),
\end{equation}
\begin{equation}
{f_{\varphi}}(x,y) = {F_{ - \alpha , - \beta }}(\exp (i2\pi {\varphi _\alpha }(u) + i2\pi {\varphi _\beta }(v))),
\end{equation}
where $F_{ - \alpha , - \beta }$ is the inverse 2D-FRFT transform with rotation angles $-\alpha$ and $-\beta$.\\\indent Set the horizontal and vertical frequency shifts as $\delta$ and $\varepsilon$, expressing in the forms of $ \exp (i2\pi x\delta ) $ and $\exp (i2\pi y\varepsilon ) $ in 2D-FRFT. The frequency shift operation ${F_{\alpha ,\beta }}^\sim(u,v)$ in 2D-FRFT is in the form of:
\begin{equation}
{F_{\alpha ,\beta }}^\sim(u,v) = \int\limits_{ - \infty }^{ + \infty } {\int\limits_{ - \infty }^{ + \infty } {{\Gamma _{\alpha ,\beta }}\cdot\exp (i2\pi x\delta + i2\pi y\varepsilon )dx} } dy,
\end{equation}
where ${\Gamma _{\alpha ,\beta }} = {K_\alpha }(x,u)\cdot{K_\beta }(y,v)\cdot f(x,y)$. Then, equation (10) can be equivalently written as follows
\begin{equation}
 {F_{\alpha ,\beta }}^\sim(u,v)  = {A_{\alpha ,\beta }}^\sim(u,v)\cdot\exp (i2\pi {\varphi _\alpha }^\sim(u) + i2\pi {\varphi _\beta }^\sim(v)),
\end{equation}
where ${A_{\alpha ,\beta }}^ \sim (u,v)$ and $\exp (i2\pi {\varphi _\alpha }^ \sim (u)+ i2\pi {\varphi _\beta }^ \sim (v))$ represent the amplitude component and phase component of equation (10).\\\indent ${A_{\alpha ,\beta }}^ \sim (u,v)$ is equivalently expressed as follows
\begin{equation}
{A_{\alpha ,\beta }}^\sim(u,v) = \left| {{F_{\alpha ,\beta }}^\sim(u,v)} \right|.
\end{equation}
Using an algebraic operation, equation (12) is further written as seen below
\begin{equation}
\begin{array}{l}
 {A_{\alpha ,\beta }}^\sim(u,v) = \left| {\int\limits_{ - \infty }^{ + \infty } {\int\limits_{ - \infty }^{ + \infty } {{\Gamma _{\alpha ,\beta }}\cdot\exp (i2\pi x\delta  + i2\pi y\varepsilon )dx} } dy} \right| \\
  = {A_{\alpha ,\beta }}(u - 2\pi \delta \sin \alpha ,v - 2\pi \varepsilon \sin \beta ). \\
 \end{array}
\end{equation}
The derivation of equation (13) is given in Appendix \emph{A} of the supporting document.\\\indent
Based on the separability of 2D-FRFT and algebraic operation, the phase parts $\exp (i2\pi {\varphi _\alpha }^ \sim (u))$ and $\exp(i2\pi {\varphi _\beta }^ \sim (v))$ in equation (11) are given as follows
\begin{small}
\begin{equation}
\begin{array}{l}
 \exp (i2\pi {\varphi _\alpha }^\sim(u)) \\
  = \exp \left\{ {i2\pi [{\varphi _\alpha }(u - 2\pi \delta \sin \alpha ) + (u\delta \cos \alpha )] - i\pi \sin \alpha \cos \alpha (2\pi ){\delta ^2}} \right\}, \\
 \end{array}
\end{equation}
\end{small}
\begin{small}
\begin{equation}
\begin{array}{l}
 \exp (i2\pi {\varphi _\beta }^\sim(v)) \\
  = \exp \left\{ {i2\pi [{\varphi _\beta }(v - 2\pi \varepsilon \sin \beta ) + (v\varepsilon \cos \beta )] - i\pi \sin \beta \cos \beta (2\pi ){\varepsilon ^2}} \right\}. \\
 \end{array}
\end{equation}
\end{small}
The derivation of equations (14) and (15) is given in Appendix \emph{B} of the supporting document.\\\indent
Let $\rho$ and $\lambda$ be horizontal and vertical spatial shifts of $f(x,y)$. The spatial shift ${F_{\alpha ,\beta }}^{'}(u,v)$ in 2D-FRFT possesses the following relation [20]:
\begin{equation}
{F_{\alpha ,\beta }}^{'}(u,v) = \int\limits_{ - \infty }^{ + \infty } {\int\limits_{ - \infty }^{ + \infty } {{\kappa _{\alpha ,\beta }}(x,y,u,v) \cdot f(x - \rho ,y - \lambda )dxdy} },
\end{equation}
where ${\kappa _{\alpha ,\beta }}(x,y,u,v) = {K_\alpha }(x,u)\cdot{K_\beta }(y,v)$. Equation (16) is equivalently given in the form of amplitude ${A_{\alpha ,\beta }}^{'}(u,v)$ and phase $\exp (i2\pi {\varphi ^{'}}_\alpha (u) + i2\pi {\varphi ^{'}}_\beta (v))$ in equation (17),
\begin{equation}
\begin{array}{l}
 {F_{\alpha ,\beta }}^{'}(u,v) = {A_{\alpha ,\beta }}^{'}(u,v) \cdot \exp (i2\pi {\varphi ^{'}}_\alpha (u) + i2\pi {\varphi ^{'}}_\beta (v)). \\
 \end{array}
\end{equation}
Using an algebraic method, ${A_{\alpha ,\beta }}^{'}(u,v)$ is equivalently written as follows
\begin{equation}
\begin{array}{l}
 {A_{\alpha ,\beta }}^{'}(u,v) = \left| {\int\limits_{ - \infty }^{ + \infty } {\int\limits_{ - \infty }^{ + \infty } {{\kappa _{\alpha ,\beta }}(x,y,u,v)\cdot f(x - \rho ,y - \lambda )dxdy} } } \right| \\
  = {A_{\alpha ,\beta }}(u - \rho \cos \alpha ,v - \lambda \cos \beta ). \\
 \end{array}
\end{equation}
The derivation of equation (18) is given in Appendix \emph{C} of the supporting document.\\\indent Since 2D-FRFT satisfies the inversed transform in equation (5), implementing the inverse 2D-FRFT w. r. t. $-\alpha$ and $-\beta$ representing the magnitude of the reconstructed amplitude component ${f_{A}}^\sim(x,y)$ from ${A_{\alpha ,\beta }}^ \sim (u,v)$ leads to the following expression,
\begin{equation}
 \left| {{f_{{A^{^\sim}}}}\left( {x,y} \right)} \right| = \left| {{F_{ - \alpha , - \beta }}({A_{\alpha ,\beta }}^\sim(u,v))} \right|.
\end{equation}
By using equation (13), equation (19) is rewritten as follows
\begin{equation}
\begin{array}{l}
 \left| {{f_A}^\sim(x,y)} \right| \\
  = \left| {{F_{ - \alpha , - \beta }}({A_{\alpha ,\beta }}^\sim(u,v))} \right| \\
  = \left| {{F_{ - \alpha , - \beta }}({A_{\alpha ,\beta }}(u - 2\pi \delta \sin \alpha ,v - 2\pi \varepsilon \sin \beta ))} \right|. \\
 \end{array}
\end{equation}
Based on equations (8) and (18), $| {{f_A}^\sim(x,y)}| $ is further expressed as
\begin{equation}
\begin{array}{l}
 \left| {{f_A}^\sim(x,y)} \right| \\
  = \left| {{F_{ - \alpha , - \beta }}({A_{\alpha ,\beta }}(u - 2\pi \delta \sin \alpha ,v - 2\pi \varepsilon \sin \beta ))} \right| \\
  = \left| {{f_A}\left( {x - 2\pi \delta \sin \alpha \cos \alpha ,y - 2\pi \varepsilon \sin \beta \cos \beta } \right)} \right| \\
  = \left| {{f_A}\left( {x - \pi \delta \sin 2\alpha ,y - \pi \varepsilon \sin 2\beta } \right)} \right|. \\
 \end{array}
\end{equation}
Similarly, the magnitude of the reconstructed phase component ${{f_{{\varphi ^\sim}}}(x,y)}$ from $\exp (i2\pi {\varphi _\alpha }^ \sim (u)+ i2\pi {\varphi _\beta }^ \sim (v))$ yields the following expression,
\begin{equation}
\begin{array}{l}
 \left| {{f_{{\varphi ^\sim}}}(x,y)} \right| \\
  = \left| {{F_{ - \alpha , - \beta }}(\exp (i2\pi {\varphi _\alpha }^\sim(u) + i2\pi {\varphi _\beta }^\sim(v)))} \right| \\
  = \left| {{F_{ - \alpha , - \beta }}(\exp (i2\pi {\varphi _\alpha }^\sim(u)) \cdot \exp (i2\pi {\varphi _\beta }^\sim(v)))} \right|. \\
 \end{array}
\end{equation}
Since 2D-FRFT is equivalent to apply FRFT on the two variables successively, mathematical manipulation of (22) yields
\begin{equation}
\begin{array}{l}
 \left| {{f_{{\varphi ^\sim}}}(x,y)} \right| \\
  = \left| {{F_{ - \alpha , - \beta }}(\exp (i2\pi {\varphi _\alpha }^\sim(u)) \cdot \exp (i2\pi {\varphi _\beta }^\sim(v)))} \right| \\
  = \left| {{F_{ - \alpha }}(\exp (i2\pi {\varphi _\alpha }^\sim(u))) \cdot {F_{ - \beta }}(\exp (i2\pi {\varphi _\beta }^\sim(v)))} \right|. \\
 \end{array}
\end{equation}
From equation (23), $|{{F_{ - \alpha }}(\exp (i2\pi {\varphi _\alpha }^\sim(u)))}|$ can be rewritten as follows
\begin{equation}
\begin{array}{l}
 \left| {{F_{ - \alpha }}(\exp (i2\pi {\varphi _\alpha }^\sim(u)))} \right| \\
  = \left| {{F_{ - \alpha }}(\exp \left\{ {i2\pi [{\varphi _\alpha }(u - 2\pi \delta \sin \alpha ) + (u\delta \cos \alpha )]} \right\})} \right| \\
  = \left| {{F_{ - \alpha }}(\exp (i2\pi {\varphi _\alpha }(u - 2\pi \delta \sin \alpha )) \cdot \exp (i2\pi u\delta \cos \alpha ))} \right|. \\
 \end{array}
\end{equation}
By definition,
\begin{equation}
{f_\varphi }(x) = {F_{ - \alpha }}(\exp (i2\pi {\varphi _\alpha }(u))).
\end{equation}
According to the separability of 2D-FRFT and equation (18), $| {{F_{ - \alpha }}(\exp (i2\pi {\varphi _\alpha }(u - 2\pi \delta \sin \alpha )))} | $ is further written as
\begin{equation}
\begin{array}{l}
 \left| {{F_{ - \alpha }}(\exp (i2\pi {\varphi _\alpha }(u - 2\pi \delta \sin \alpha )))} \right| \\
  = \left| {{f_\varphi }(x - 2\pi \delta \sin \alpha \cos ( - \alpha ))} \right| \\
  = \left| {{f_\varphi }(x - 2\pi \delta \sin \alpha \cos \alpha )} \right|. \\
 \end{array}
\end{equation}
Employing equation (13) and substituting (26) into (24) yields equation (27)
\begin{equation}
\begin{array}{l}
 \left| {{F_{ - \alpha }}(\exp (i2\pi {\varphi _\alpha }^\sim(u)))} \right| \\
  = \left| {{F_{ - \alpha }}(\exp (i2\pi {\varphi _\alpha }(u - 2\pi \delta \sin \alpha )) \cdot \exp (i2\pi u\delta \cos \alpha ))} \right| \\
  = \left| {{f_\varphi }(x - 2\pi \delta \sin \alpha \cos \alpha  - 2\pi \delta \cos \alpha \sin ( - \alpha ))} \right| \\
  = \left| {{f_\varphi }(x)} \right|. \\
 \end{array}
\end{equation}
The equivalence between $|{{F_{ - \beta }}(\exp (i2\pi {\varphi _\beta }^\sim(v)))}|$ and $|{{f_\varphi }(y)}|$ can be similarly verified, thus
\begin{equation}
\begin{array}{l}
 \left| {{F_{ - \beta }}(\exp (i2\pi {\varphi _\beta }^\sim(v)))} \right| \\
  = \left| {{F_{ - \beta }}(\exp (i2\pi {\varphi _\beta }(v - 2\pi \varepsilon \sin \beta ))\exp (i2\pi v\varepsilon \cos \beta ))} \right| \\
  = \left| {{F_{ - \beta }}(\exp (i2\pi {\varphi _\beta }(v)))} \right| \\
  = \left| {{f_\varphi }(y)} \right|.\\
 \end{array}
\end{equation}
Substituting equations (27) and (28) into (22) yields,
\begin{equation}
\begin{array}{l}
  \left| {{f_{{\varphi ^\sim}}}(x,y)} \right| \\
 =\left| {{F_{ - \alpha , - \beta }}(\exp (i2\pi {\varphi _\alpha }^\sim(u) + i2\pi {\varphi _\beta }^\sim(v)))} \right| \\
  = \left| {{F_{ - \alpha }}(\exp (i2\pi {\varphi _\alpha }(u)))} \cdot {{F_{ - \beta }}(\exp (i2\pi {\varphi _\beta }(v)))} \right| \\
  = \left|{F_{ - \alpha , - \beta }}(\exp (i2\pi {\varphi _\alpha }(u) + i2\pi {\varphi _\beta }(v)))\right|\\
  = \left| {{f_{\varphi}}(x,y)} \right|. \\
 \end{array}
\end{equation}
Equations (21) and (29) demonstrate that the magnitude of reconstruction from amplitude-only information in 2D-FRFT domain is due to the corresponding shift with respect to the frequency shift operations. Nevertheless, the magnitude of reconstruction from phase-only information in 2D-FRFT domain has no shift at all.\\\indent Moreover, since $x$ and $y$ are integers in the field of digital image processing, $ \exp (i2\pi x\delta ) $ and $\exp (i2\pi y\varepsilon ) $ change into periodic functions and the period is 1 for $\delta$ and $\varepsilon$, respectively. Therefore, during the following computer simulations, $\delta$ and $\varepsilon$ satisfy the relation (30)
\begin{equation}
\left\{ \begin{array}{l}
 \exp (i2\pi x\delta ) = \exp (i2\pi x(\delta  + 1)), \\
 \exp (i2\pi x\varepsilon ) = \exp (i2\pi x(\varepsilon  + 1)). \\
 \end{array} \right.
\end{equation}
\subsection{Simulations}
In this subsection, the impact of frequency shift on the amplitude and phase components in 2D-FRFT is shown by the following computer simulations. During the simulations, the rotation angles are selected as $\alpha$=$\beta$=36$^o$ and frequency shift parameters are set as ($\delta$ = 0.2, $\varepsilon$ = 0) and ($\delta$ = 10.2, $\varepsilon$ = 0) randomly. The simulation results on image `Lena' are illustrated in Fig. 1. From the simulation results, it is observed the phase information is frequency shift-invariant for image reconstruction in 2D-FRFT domain while the amplitude information does not possess this property. Moreover, experimental results on the periodic characteristics of $ \exp (i2\pi x\delta )  (\delta =0.2, \delta =10.2)$ in 2D-FRFT domain ($\alpha$=$\beta$=36$^o$) are shown in Fig. 1 (d) to Fig. 1 (i).
\section{Applications}
As two degrees of freedom are provided in 2D-FRFT, raising the potential to generate more security [24], 2D-FRFT has been widely applied in the field of image encryption. In this section, we present utilization of the property of frequency shift in 2D-FRFT, which is expected to find applications in the aforementioned fields to improve the robustness.\\\indent Information processing in the encrypted domain has attracted considerable research interests [25-26]. In [27], a double random phase fractional order Fourier domain encoding scheme is proposed for image encryption to enhance the level of security. It demonstrated that the double random phase method is robust against attacks such as occlusion, crop, and so forth [27]. However, there is a chronic issue [28-29] that frequency shift can introduce interference into phase information and decrease the robustness of the double random phase encoding scheme. Since the image reconstruction from phase information satisfies the frequency shift-invariance property, it has potential to extract encryption information/data even when frequency shift attacks exist. In the following experiments, we will select the method of double random phase encoding used in [27] to demonstrate the effectiveness of the frequency shift-invariant property in image encryption.\\\indent In the double random phase encoding method, an independent random function $r(x,y)$ is uniformly distributed in the interval [0 2$\pi$] and the rotation angles are set as $\alpha$=$\beta$=9$^o$ randomly. Then, the method of random phase encoding on a two dimensional signal \emph{$I(x,y)$} is written as follows
\begin{equation}
g(\varsigma ,\eta ) = \int {\int {I(x,y)\cdot exp(2\pi i \cdot r(x,y))\cdot {\bf{\Phi }} \cdot dxdy} },
\end{equation}
\begin{equation}
i=(-1)^{1/2},
\end{equation}
where ${\bf{\Phi }}={K_{\alpha  = {9^o}^,\beta  = {9^o}}}(\varsigma  ,\eta , x , y )$ is the transform kernel function in 2D-FRFT, and the function $g(\varsigma  ,\eta )$ is the encrypted signal.\\\indent Since it satisfies the property of inverse in 2D-FRFT domain, the original signal \emph{$I(x,y)$} can be recovered with the correct independent random functions and rotation angles. Nevertheless, when the frequency shifts $exp(i 2\pi x \delta )$ and $ exp(i 2\pi y \varepsilon)$ are introduced, $\emph{I(x,y)}* exp(i 2\pi x \delta ) * exp(i 2\pi y \varepsilon)* exp(2\pi  * i * r(x,y))$ will replace $\emph{I(x,y)} * exp(2\pi  * i * r(x,y))$ in equation (31), resulting in failures of the encrypted information/data recovery even with the correct independent random functions and rotation angles. However, we can recover the encryption information/data successfully benefiting from the property of frequency shift-invariance from phase information in 2D-FRFT domain.\\\indent Experiments are provided to demonstrate the effectiveness of this property against the frequency shift attack with rotation angles ($\alpha$=$\beta$=9$^o$) in 2D-FRFT domain shown in Fig. 2. In Fig. 2, when there is no frequency shift in Fig. 2(b), the key image is successfully recovered straightforwardly in Fig. 2(c). When there is frequency shift existing in Fig. 2(d), we see the failure without using the frequency shift-invariant property shown in Fig. 2(e), and the success using the property shown in Fig. 2(f). \\\
\centerline{\includegraphics[height=4.2in,width=3.6in]{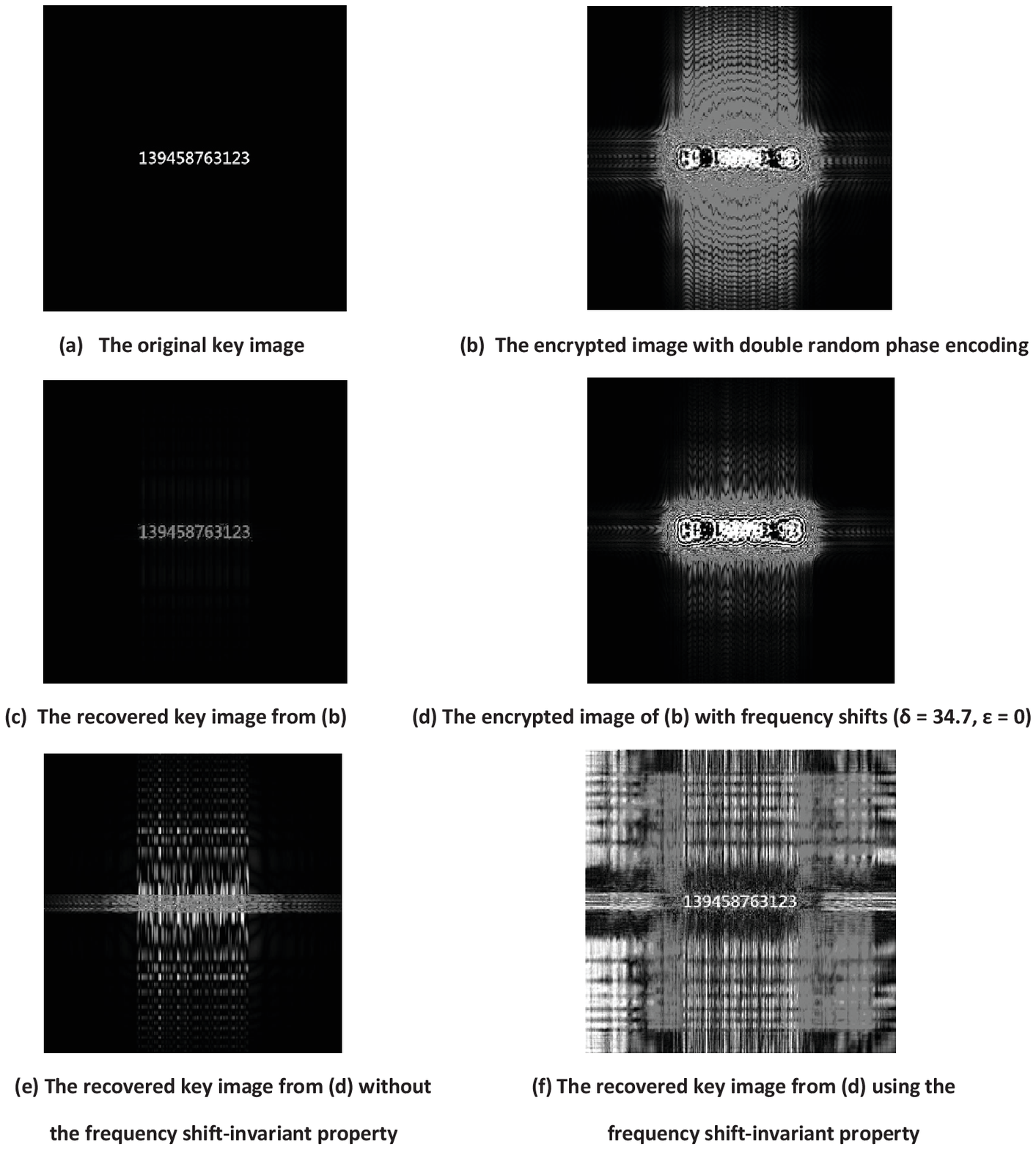}}\\ { {Fig. 2 Experiments on image encryption with the frequency shift-invariance property}}
\section{Conclusions}
In this letter, the property of frequency shift operation, from the amplitude and phase information in 2D-FRFT domain, has been studied. It is demonstrated that the magnitude of image reconstruction from phase information is frequency shift-invariant while the property does not hold for the amplitude information. Experiments are provided, illustrating the effectiveness of the property in improving the robustness of image encryption.




\ifCLASSOPTIONcaptionsoff
  \newpage
\fi

\end{document}